# Investigation of guidelines for improving spatial resolution in direct-modulation BOCDR

Seiga Ochi,[1] Kouta Ozaki,[1] Kohei Noda,[1,2,3] Heeyoung Lee,[4] Kentaro Nakamura,[2] and Yosuke Mizuno[1]

[1] *Faculty of Engineering, Yokohama National University, 79-5 Tokiwadai, Hodogaya-ku, Yokohama 240-8501, Japan*

[2] *Institute of Innovative Research, Tokyo Institute of Technology, 4259, Nagatsuta-cho, Midori-ku, Yokohama 226-8503, Japan*

[3] *Graduate School of Engineering, The University of Tokyo, 7-3-1 Hongo, Bunkyo-ku, Tokyo 113-8656, Japan*

[4] *Graduate School of Engineering and Science, Shibaura Institute of Technology, 3-7-5 Toyosu, Koto-ku, Tokyo 135-8548, Japan*

*Author e-mail addresses: ochi-eaiga-xv@ynu.jp, mizuno-yosuke-rg@ynu.ac.jp*

**Abstract:** The spatial resolution of direct-modulation Brillouin optical correlation-domain reflectometry is studied with respect to modulation amplitude and frequency. Results suggest that optimal resolution improvement is achieved by increasing modulation amplitude first, followed by frequency.

## 1. Introduction

The deterioration of social infrastructure due to aging or damage caused by disasters has emerged as a significant problem. Fiber optic sensors, which offer many advantages not found in electric sensors, have been drawing attention. A variety of methods using optical fibers have been developed for distributed measurements, depending on the type of scattered light to be analyzed and the analytical method to be used. Brillouin scattering exhibits characteristics dependent on both strain and temperature, and a considerable amount of research has been conducted to realize distributed strain and temperature sensors using this phenomenon. Representative distributed measurement methods using Brillouin scattering include time-domain methods such as Brillouin optical time-domain reflectometry (BOTDR) [1] and analysis (BOTDA) [2], frequency-domain methods such as Brillouin optical frequency-domain reflectometry (BOFDR) and analysis (BOFDA) [3], and correlation-domain methods such as Brillouin optical correlation-domain reflectometry (BOCDR) [4] and analysis (BOCDA) [5]. Herein, we focus on BOCDR, which can measure the distributions of strain and temperature with relatively high spatial resolution by injecting light into only one end of a fiber under test (FUT).

Figure 1 shows the standard configuration of BOCDR. By directly modulating the drive current of the laser, frequency modulation is imposed on the output light. As a result, a correlation peak (measurement point) is generated in the FUT, enabling distributed measurements to be performed by sweeping this peak [4]. The spatial resolution of BOCDR $\Delta z$ is known to be inversely proportional to the product of two modulation parameters: modulation amplitude $\Delta f$ and modulation frequency $f_m$ [6]. Therefore, theoretically speaking, increasing either of these modulation parameters will improve spatial resolution equally, but this is not always the case in practice. In the direct-modulation scheme of BOCDR, which modulates the output light frequency by directly modulating the drive current of the laser, power modulation is inevitably accompanied by frequency modulation of the output light. The magnitude of this power modulation increases with increasing modulation amplitude $\Delta f$, but remains almost constant with increasing modulation frequency $f_m$. Therefore, the contribution of these two modulation parameters to spatial resolution should be different (e.g., when $f_m$ is doubled, $\Delta z$ is theoretically halved, but when *Δf* is doubled, *Δz* may not be halved as expected). Another performance metric is measurement range $d_m$ (measurable distance), which is inversely proportional to the modulation frequency $f_m$. Therefore, as spatial resolution and measurement range are in a trade-off relationship with respect to $f_m$, it cannot be concluded simply that increasing modulation frequency is the best way to improve spatial resolution. To date, regarding the contribution of both the modulation parameters to the spatial resolution, experimental verification has not yet been reported.

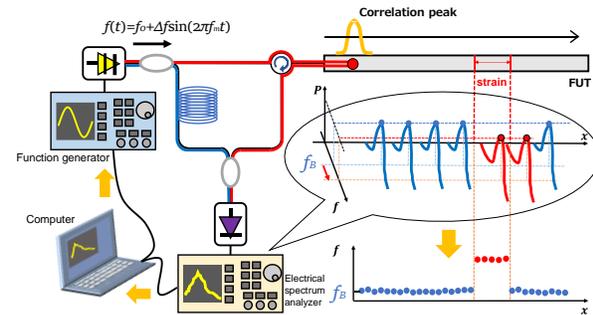

**Fig. 1** Schematic setup of direct-modulation BOCDR.

In this work, we aim to qualitatively compare the contributions of modulation amplitude *Δf* and modulation frequency $f_m$ to spatial resolution, and to provide guidelines for improving the spatial resolution of direct-modulation BOCDR by controlling modulation parameters.

## 2. Principles

In a standard BOCDR system (Fig. 1), frequency modulation is achieved by directly modulating the laser driving current, resulting in an output light with frequency modulation. The frequency of the output light can be described by

$$f(t) = f_0 + \Delta f \sin(2\pi f_m t), \tag{1}$$

where $f_0$ is the central frequency. As a result of frequency modulation, a correlation peak (measurement point) is generated in the FUT, and scanning this peak enables distributed measurement to be carried out [4]. The spatial resolution of BOCDR is theoretically known to be inversely proportional to the product of modulation amplitude and modulation frequency [6]. However, for the direct modulation method that inevitably accompanies power modulation, the contribution of modulation amplitude and modulation frequency to spatial resolution should be different.

Furthermore, the theoretical spatial resolution Δz of BOCDR is given by [6]

$$\Delta z \approx \frac{c \Delta \nu_B}{2\pi n f_m \Delta f}, \tag{2}$$

where *c* is the speed of light, $\Delta \nu_B$ is the Brillouin linewidth (~30 MHz in silica fibers), and *n* is the effective refractive index of the core (~1.46 in silica fibers). Thus, the theoretical spatial resolution is inversely proportional to both modulation amplitude *Δf* and modulation frequency $f_m$, and increasing either *Δf* or $f_m$ will improve the spatial resolution (i.e., the value becomes smaller). However, in the direct-modulation method of BOCDR, which inevitably involves power modulation (Fig. 2), the actual spatial resolution obtained does not necessarily coincide with Eq. (2) (due to the effect of apodization). Simply put, the magnitude of power modulation does not change even when $f_m$ is increased while maintaining *Δf*. On the other hand, increasing *Δf* while maintaining $f_m$ results in an increase in the magnitude of power modulation. Therefore, it is anticipated that the contribution to spatial resolution is different between the two modulation parameters.

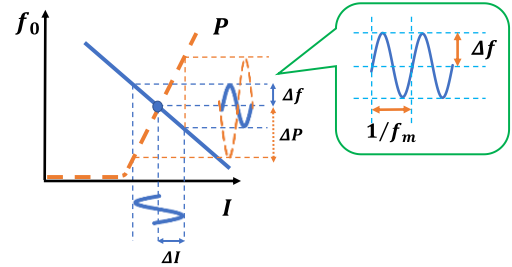

**Fig. 2** Relationship between laser driving current and output frequency/power.

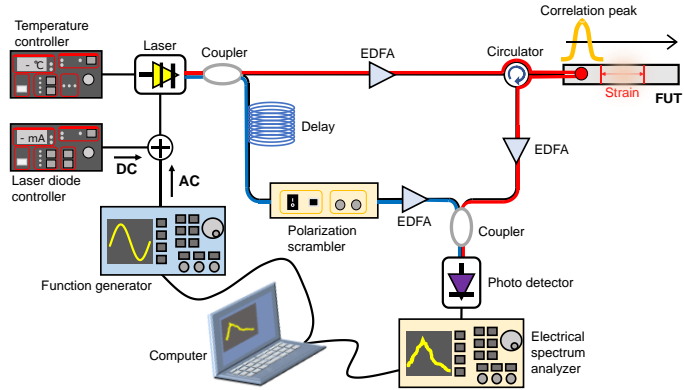

**Fig. 3** Experimental setup of BOCDR with direct modulation. EDFA: erbium-doped fiber amplifier.

## 3. Experiments and Discussion

We employed the experimental setup

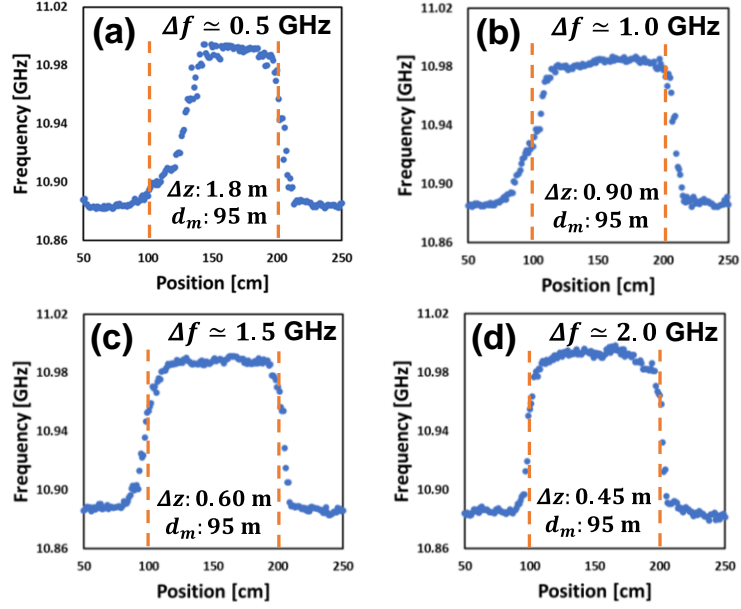

**Figs. 4** Measured BFS distributions ($f_m$: constant, $\Delta f$: changed).

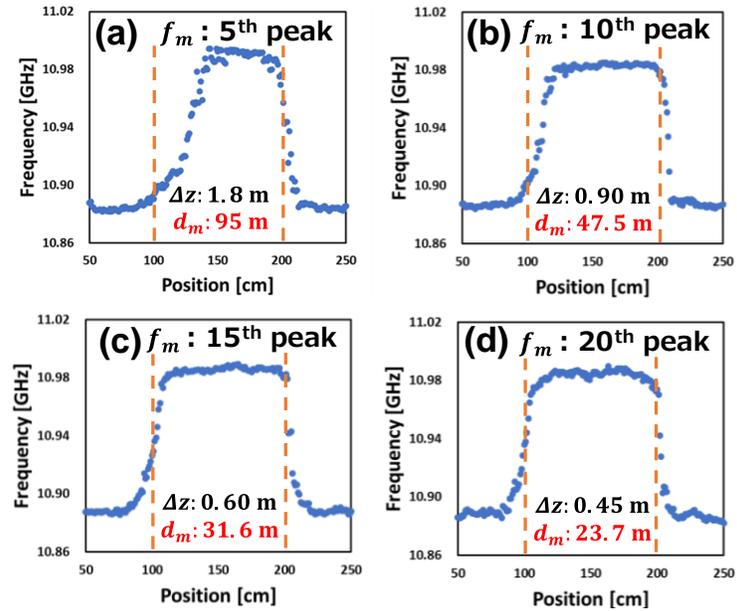

**Figs. 5** Measured BFS distributions ($f_m$: changed, $\Delta f$: constant).

depicted in Fig. 3. First, a 100-cm-long section 100–200 cm away from the front of the 4.5 m-long FUT was subjected to approximately 0.3% strain. When the modulation frequency $f_m$ was swept from 214.30 kHz to 215.30 kHz, the location of the 1st-order correlation peak shifted from 50 cm to 250 cm from the front of the FUT. For instance, when the value of $f_m$ was multiplied by 5 (ranging from 1071.5 kHz to 1076.5 kHz), the 5th-order correlation peak shifted within that range. In this experiment, we used this 200-cm-long section as the measured range and compared the improvement in strain detection performance for the 100-cm-long strain applied within that range, by increasing either the modulation amplitude $\Delta f$ or the modulation frequency $f_m$, as $\Delta z$ is inversely proportional to $\Delta f$ and $f_m$.

Initially, we performed a distributed measurement with modulation parameters set to their initial values of "$\Delta f$ = 0.50 GHz, $f_m$: 5th peak". Then, we fixed one of the modulation parameters and conducted distributed measurements by setting the other modulation parameter to 2, 3, and 4 times the initial value. We repeated this procedure with both modulation parameters and compared the results to assess the improvement in measurement performance. The theoretical spatial resolution at the initial values was 1.8 m, and the measurement range was approximately 96 m. Thus, if either of $\Delta f$ or $f_m$ was doubled, the theoretical spatial resolution fell below the length of the actual strained range.

The results of the distributed BFS measurement near the strained range are shown in Figs. 4 and 5. Figures 4(a)–(d) show the results when $\Delta f$ is increased, and Figs. 5(a)–(d) show the results when $f_m$ is increased. In (a)–(d) of both figures, the horizontal axis represents the position from the front of the FUT, and the orange dotted lines indicate the actual strained range (between 100 and 200 cm). In Figs. 4 and 5, (b)–(d) have equivalent theoretical spatial resolution for both modulation parameters. There was no significant difference observed in the distributed BFS measurement results when either $\Delta f$ or $f_m$ was increased. Thus, it became clear that the influence of power modulation due to direct modulation of the laser is extremely small under these experimental conditions. On the other hand, the measurement range $d_m$ is inversely proportional to the modulation frequency $f_m$ only, so Fig. 4(d) has a 4 times longer measurement range than that of Fig. 5(d). Therefore, considering the measurement range, increasing $\Delta f$ is preferable to increasing $f_m$. However, it is theoretically known that the upper limit of the modulation amplitude is restricted to half of the BFS due to the influence of noise caused by Rayleigh scattering [4]; in reality, the upper limit can also be determined by the load on the laser. Therefore, to improve the spatial resolution of BOCDR based on direct modulation of the laser, by controlling the modulation parameters, the guideline should be to "first increase the modulation amplitude to the upper limit, and then increase the modulation frequency."

## 5. Conclusions

In this study, we compared the improvement in spatial resolution of BOCDR using direct modulation when only one of the two modulation parameters (modulation amplitude and frequency) was changed and investigated the contribution of each parameter to spatial resolution. From the experimental results, it was clarified that the influence of power modulation due to direct modulation is extremely small when improving spatial resolution by changing modulation amplitude. Since the measurement range is inversely proportional to modulation frequency only, it is preferable to increase modulation amplitude rather than modulation frequency when improving spatial resolution, taking this into consideration as well. Therefore, we propose the guideline that when improving the spatial resolution of BOCDR based on direct modulation of the laser, modulation amplitude should be increased to the upper limit (half of the BFS) first and then modulation frequency should be increased.

**Acknowledgements**
This work was partially supported by the Japan Society for the Promotion of Science (JSPS) KAKENHI (Grant Nos. 21H04555, 22K14272, and 20J22160).